\def\year{2016}\relax
\documentclass[letterpaper]{article}
\usepackage{aaai16}
\usepackage{times}
\usepackage{helvet}
\usepackage{courier}
\usepackage{graphicx}
\usepackage{caption}

\begin{document}
%
\title{A Semi-automatic Method for Efficient Detection of Stories on Social Media}

\author{Soroush Vosoughi, Deb Roy\\
Massachusetts Institute of Technology\\ Cambridge, MA 02139\\ \tt{soroush@mit.edu, dkroy@media.mit.edu}}


\maketitle
\begin{abstract}
\begin{quote}
Twitter has become one of the main sources of news for many people. As real-world events and emergencies unfold, Twitter is abuzz with hundreds of thousands of stories about the events. Some of these stories are harmless, while others could potentially be life-saving or sources of malicious rumors. Thus, it is critically important to be able to efficiently track stories that spread on Twitter during these events. In this paper, we present a novel semi-automatic tool that enables users to efficiently identify and track stories about real-world events on Twitter. We ran a user study with 25 participants, demonstrating that compared to more conventional methods, our tool can increase the speed and the accuracy with which users can track stories about real-world events.
\end{quote}
\end{abstract}

\setlength{\belowcaptionskip}{-7pt}

\section{Introduction}
In recent years, Twitter, a social media platform with hundreds of millions of users, has become a major source of news for people \cite{stassen2010your}. This is especially true for breaking-news about real-world events \cite{kwak2010twitter}. The 2011 Japanese earthquake, the 2013 Boston marathon bombings, and the 2015 Paris shootings are just three examples of an event where Twitter played major role in the dissemination of information. However, given the great volume of tweets generated during these events, it becomes extremely difficult to make sense of all the information that is being shared. In this paper, we present a semi-automatic tool that combines state-of-the-art natural language processing and clustering algorithms in a novel way, enabling users to efficiently and accurately identify and track stories that spread on Twitter about particular events. The output of our system can also be used by rumor verification systems to substantiate the veracity of rumors on Twitter \cite{vosoughi2015automatic}.

A lot of the messages that are posted on Twitter about events are not directly related to any stories about the event itself. For instance, a tweet talking about how scared someone is about an event, does not contain useful information about the event itself. Tweets that are related to a story usually contain \emph{assertions}. Therefore, our method focused on first identifying assertions about events. These assertions could be anything from eye-witness testimony, to false rumors, or reports from the media or law enforcement agencies. 





\subsection{What is an Assertion?}
An assertion is an utterance that commits the speaker to the truth of the expressed proposition. For example, the tweet, "there is a third bomber on the roof" contains an assertion, while the tweet, "I hate reporters!" does not (it contains an expression). 
It has been shown than more than half of tweets about events do not contain assertions \cite{vosoughi_act_2016}. Thus, by filtering all non-assertions tweets, we can drastically reduce the number of tweets that need to be analysed for story detection.




\subsection{System Overview}
\begin{figure}[h]
\centering
\includegraphics[width=1.\columnwidth]{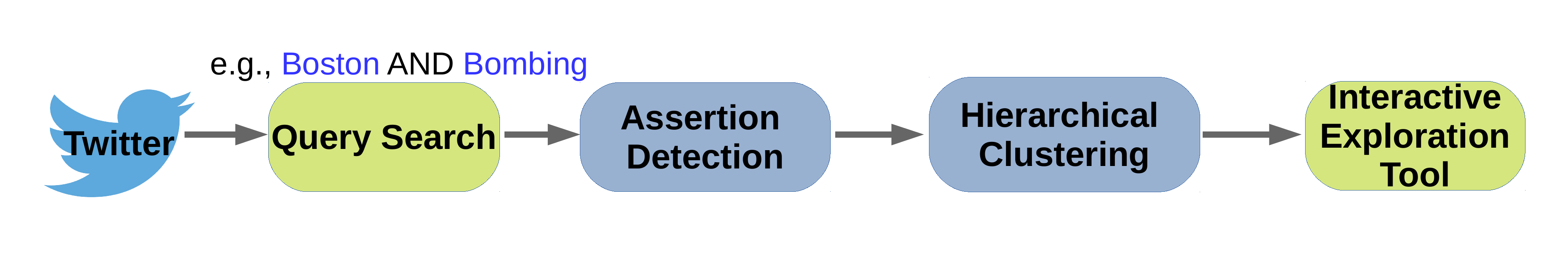}
\caption{An overview of the system pipeline.}
\label{fig:pipeline}
\end{figure}

An overview of the system can be seen in Figure \ref{fig:pipeline}. In the figure, the modules that are fully automatic are shown in blue while the modules requiring manual input are shown in green. Currently, the system only works on Twitter, though we plan to expand it to cover other publicly available social media platforms, such as Reddit.

The first module in the system is a standard boolean query search, specified by the user. The purpose of this query is to limit the scope of the data that is being analysed to one event. This query can be about anything but works best if it is about a well-defined event. For example, in this figure, the query is \emph{Boston AND Bombing}, which picks out tweets about the 2013 Boston marathon bombings. These tweets are next passed to the ``automatic'' parts of the system, an \emph{Assertion Detector} module and a \emph{Hierarchical Clustering} module. 

Raw tweets about an event feed directly into the assertion detector, which automatically filters the tweets for only those containing assertions (tweets not containing assertions are discarded at this stage). These tweets are then clustered in a hierarchical manner, based on the their semantic similarity. In theory, these clusters should mostly contain tweets that are making similar assertions. The hierarchical clusters (and their contents, including the text and the meta-data of the tweets they contain) are passed to the user-facing, interactive exploration tool. The exploration tool can be used to identify, investigate, and track stories, that are spreading about an event on Twitter.  


\section{Detecting Assertions in Tweets}
Assertions are a class of speech-acts. In order to detect assertions in tweets, a speech-act classifier is needed. We manually annotated $7,000$ tweets about several different real-world events. We labelled these tweets as either containing assertions or not. Of the $7,000$ tweets, $3,290$ ($47\%$) of those tweets containing assertions and the rest containing one of the other speech-acts. These tweets were used to train a state-of-the-art supervised Twitter speech-act classifier, developed by Vosoughi et al. \cite{vosoughi_act_2016}. 

\begin{figure}[htbp]
\centering
\includegraphics[width=.70\columnwidth]{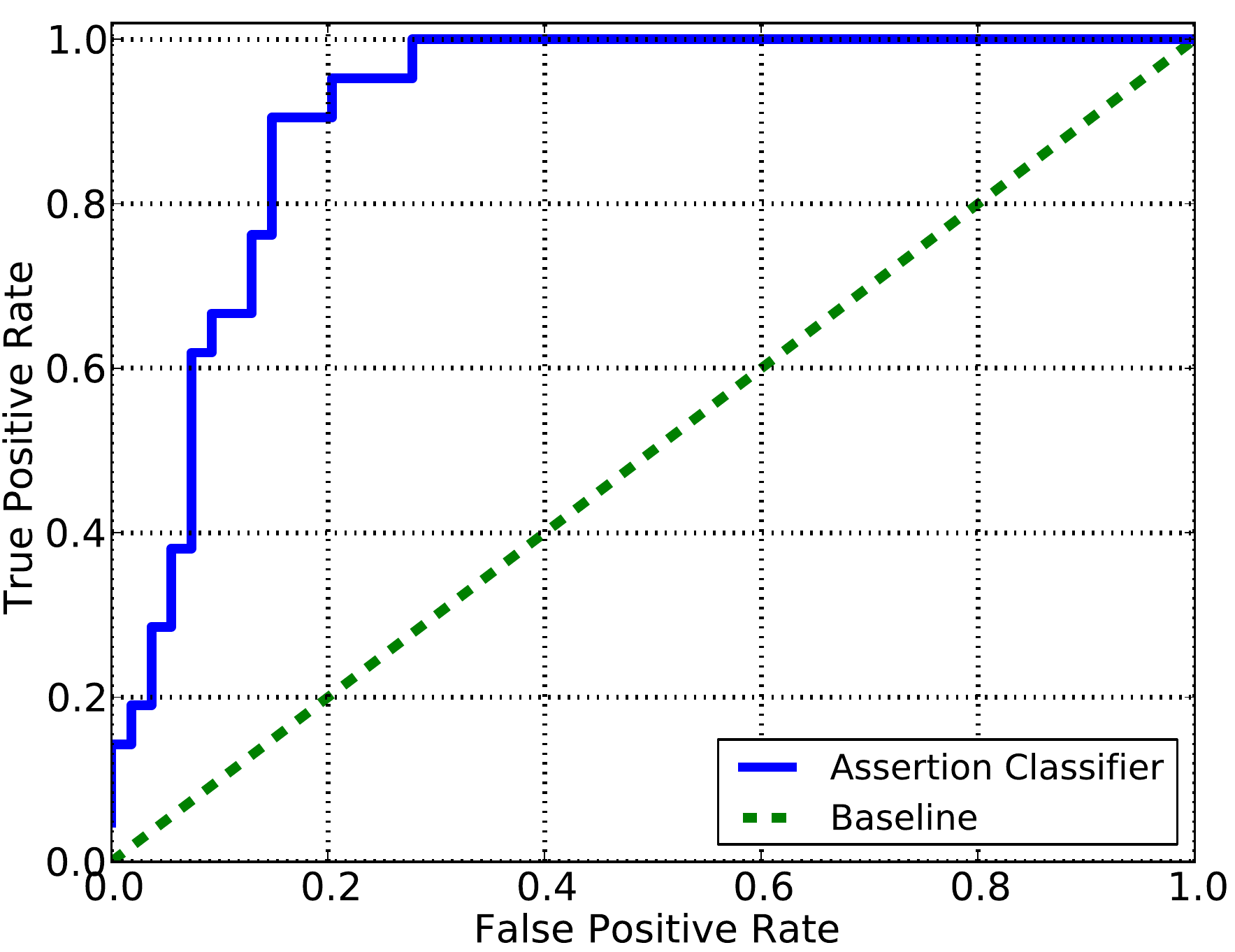}
\caption{The receiver operating characteristic (ROC) curve of the assertion classifier.}
\label{fig:assertion_roc}
\end{figure}

Since, we were interested in only detecting assertions, we turned the speech-act classifier to a binary assertion classifier (by collapsing all the non-assertion classes into one class). We evaluated the classifier using 20-fold cross validation, with the F-score for classifying assertions being $.86$. The performance of this classifier is better illustrated by its ROC curve in Figure \ref{fig:assertion_roc}.


\section{Hierarchical Clustering of Tweets}
The next part in the automatic processing pipeline is the hierarchical clustering of semantically similar tweets, in order to group together tweets making similar assertions. The output of hierarchical clustering can best be described as a dendrogram.
At the lowest level of the dendrogram, all tweets belong to their own clusters. At the very root of the tree is a single cluster, containing all the tweets. Users can explore the clusters at any level. A partition lower in the tree (further from the root) would yield more clusters, with each cluster containing fewer number of tweets. Conversely, a partition higher in the tree would yield less clusters, with each containing greater number of tweets. Depending on the event, there could be thousands of clusters at different levels. It will be up to the users to decide how to best cut-off and explore the clusters. For example, if the event in question is a very local event, meaning that there are not many tweets about the event, then perhaps a partition higher in the tree would be more useful and vice-versa.



Hierarchical clustering of text documents is a well-studied problem. However, as was the case with speech-act classification, the noisy, unconventional and most importantly short nature of the language used on Twitter, greatly reduce the performance of conventional hierarchical document clustering methods. Thus, we developed a novel hierarchical clustering method for Twitter, using very recent advances in Twitter natural language processing techniques. 

In the next sections, we will describe a conventional hierarchical clustering method, followed by our novel method. Both methods were implemented so that the performance of our novel method could be benchmarked.


\subsection{Conventional Method}
Generally speaking, there are two strategies for hierarchical clustering: 
\begin{itemize}
\item \emph{Agglomerative}: This is a "bottom up" approach; each observation starts in its own cluster, and pairs of clusters are merged as one moves up the hierarchy.
\item \emph{Divisive}: This is a "top down" approach; all observations start in one cluster, and splits are performed recursively as one moves down the hierarchy.
\end{itemize}

The complexity of agglomerative clustering is polynomial at $O(n^{3})$, while the complexity of divisive clustering is exponential at $O(2^{n})$. Given the potentially large number tweets about an event, we decided to use Hierarchical Agglomerative Clustering (HAC), given its lower complexity. 

To do any sort of clustering of documents (such as tweets), a similarity function is needed, to measure the similarity between different documents and decide which clusters to merge or divide. A standard metric used to measure similarity between text documents is \emph{TF-IDF} combined with \emph{cosine similarity}. TF-IDF is a method of converting text into numbers so that it can be represented meaningfully by a vector. TF-IDF is the product of two statistics, \emph{TF} or Term Frequency and \emph{IDF} or Inverse Document Frequency. 
Using TF-IDF, a vector for each document is derived. The set of documents in our collection is then viewed as a set of vectors in a vector space with each term  having its own axis. Similarity between two documents is measured using cosine similarity. With this similarity function, we can hierarchically cluster tweets using HAC. 

\subsection{Novel Method}
TF-IDF, combined with cosine similarity is usually a good method of measuring similarity between documents. However, tweets are short (140 characters), irregular text whose topic-level information can be hardly expressed by TF-IDF representations. An alternative method, is to use a similarity metric that is adapted to this platform. 

\subsubsection{Twitter Paraphrase Identification}
We implemented the Twitter paraphrase identification method proposed recently by Asli Eyecioglu and Bill Keller \cite{eyecioglu2015asobek} (winners of SemEval-2015 in this category) to measure similarity between pairs of tweets. 
This method is for identifying Twitter paraphrase pairs, where paraphrase identification is defined as "the task of deciding whether two given text fragments have the same meaning". This method takes a pair of tweets and makes a binary judgement on whether these two tweets are paraphrases. For example, the tweets, "Amber alert gave me a damn heart attack" and "That Amber alert scared the crap out of me" are a paraphrase pair, while the tweets "My phone is annoying me with these amber alert", and "Am I the only one who dont get Amber alert" are not a paraphrase pair.

We used a dataset called the Twitter Paraphrase Corpus (TPC) \cite{xu2014extracting} for training and testing our model. The dataset contains 18K tweet pairs 1K test data, with $35\%$ those pairs being paraphrases, and $65\%$ non-paraphrases. We trained an linear SVM classifier using the features proposed in that paper. These features are based on overlap of word level and character level n-grams. To begin, the text in each tweet is cleaned and represented as a set of tokens, where a token can be a character or word unigram or bigram. The overlap features are then created using set operations: size of the union of the tokens, size of the intersection of the tokens, and the size of the set of tokens.

Of all the combinations of overlap features, the following six features were shown to be the most informative: \emph{union of word unigrams}, \emph{union of character bigrams}, \emph{intersection of word unigrams}, \emph{intersection of character bigrams}, \emph{sizes of tweet 1 and tweet 2}. The linear SVM trained on these features achieved an F-score of $.67\%$. Other than the great performance, this method is very fitted to our use-case since both feature extraction and classification are extremely fast. Given that sometimes the number of tweets about a particular event could be in the millions, this is extremely important.

All possible pairs of tweets that make it past our assertion detector (which is ${N  \choose 2}$ pairs, N being the number of tweets containing assertions), are passed through this binary classifier, to be classified as paraphrases or not. The results are used to create an undirected graph, with each of the $N$ tweets being represented as a node, and edges between nodes representing paraphrase pairs. This graph is used to construct hierarchical clusters of tweets.

\subsubsection{Clustering using Community Detection}
Given this undirected graph of tweets, we can use efficient community detection methods, to detect communities, or "clusters" of tweets with similar meaning assertions. 

We used a method called the, \emph{Louvain} \cite{de2011generalized} for this purpose. The Louvain method is a simple and efficient method for community detection in very large networks. It uses a greedy optimization method that runs in time $O(n\log{}n)$, outperforming other community detection methods in terms of computation time, while  performing on par, if not better, than other methods when it comes to the accuracy and quality of the extracted communities \cite{aynaud2013multilevel}. It is however the speed of this method which is its main advantage, as it takes only two minutes to analyze a typical network of 2 million nodes. This is very important for applications requiring real-time clustering, such as ours. Also, crucial for our task, the Louvain method generates hierarchical structures. 

The idea behind the method is the greedy optimization of the \emph{modularity} of the graph. Modularity is defined as a value between $-1$ and $1$ that measures the density of links inside communities compared to the links between communities. The method consists of two steps. First, the method looks for "small" communities by optimizing modularity locally. Second, it aggregates nodes belonging to the same community and builds a new network whose nodes are the communities. These steps are repeated iteratively until a maximum of modularity is attained and a hierarchy of communities is produced \cite{de2011generalized}. These hierarchical communities are analogous to the hierarchical clusters generated by HAC, in that these communities contain similar assertions.

\section{Experiments}

\subsubsection{Analysis 1}
Since our system is semi-automatic, the evaluation is based on user-centric criteria: the accuracy and the speed with which users can identify assertions about a particular event on Twitter. To do this, we used \emph{snopes.com} to identify several rumors (a rumor is an unverified assertion) that had spread on Twitter for three different events: the 2013 Boston marathon bombings (10 rumors), the 2011 Japanese earthquake (6 rumors), and the 2015 Charlie Hebdo shooting in Paris (5 rumors). We manually collected between $150$ to $500$ English-language tweets for each rumor. For each event, we also mixed in several hundred random tweets about the event, not related to any of the rumors (in order to mimic a real-world situation where there is a mixture of tweets containing rumors and tweets that do not). In total there were 9,825 tweets across all three events (6,825 about one of the 21 rumors, and 3,000 random tweets about one of the three events).  

We asked a group of twenty five undergraduates to identify the rumors for each of the events. We divided the subjects into five groups. The first group used a version of our tool that only applied HAC to the data (without any assertion filtering), the second group's tool applied assertion detection (AD) and HAC to the data (this was an earlier version of our tool \cite{vosoughi_rd_2015}), the third group's tool applied the Louvain clustering method (without assertion filtering), the fourth group's tool applied AD and Louvain clustering to the data. The fifth group was the control group, their tool did not process the data at all, it put all the tweets (including non-assertions) into one giant cluster. Note that when using any of the hierarchical clustering algorithms, the users were allowed to explore the fully hierarchy. The tool did not decide on the number of clusters (i.e. where to cut the tree).

The subjects did not know beforehand how many rumors, if any, the dataset contained. Each person worked independently and was allowed five minutes per event. The time limit was set for two reasons. First, the limit was set in order to mimic a real-time emergency situation, for which our tool is designed. Second, given the relatively small size of the rumor dataset used for this experiment, given enough time, most subjects would be able to correctly identify all or most of the rumors using any of the tools, therefore a time-limit was needed to better differentiate between the different versions of the tool.

\begin{figure}[htb]
\centering
\includegraphics[width=.80\columnwidth]{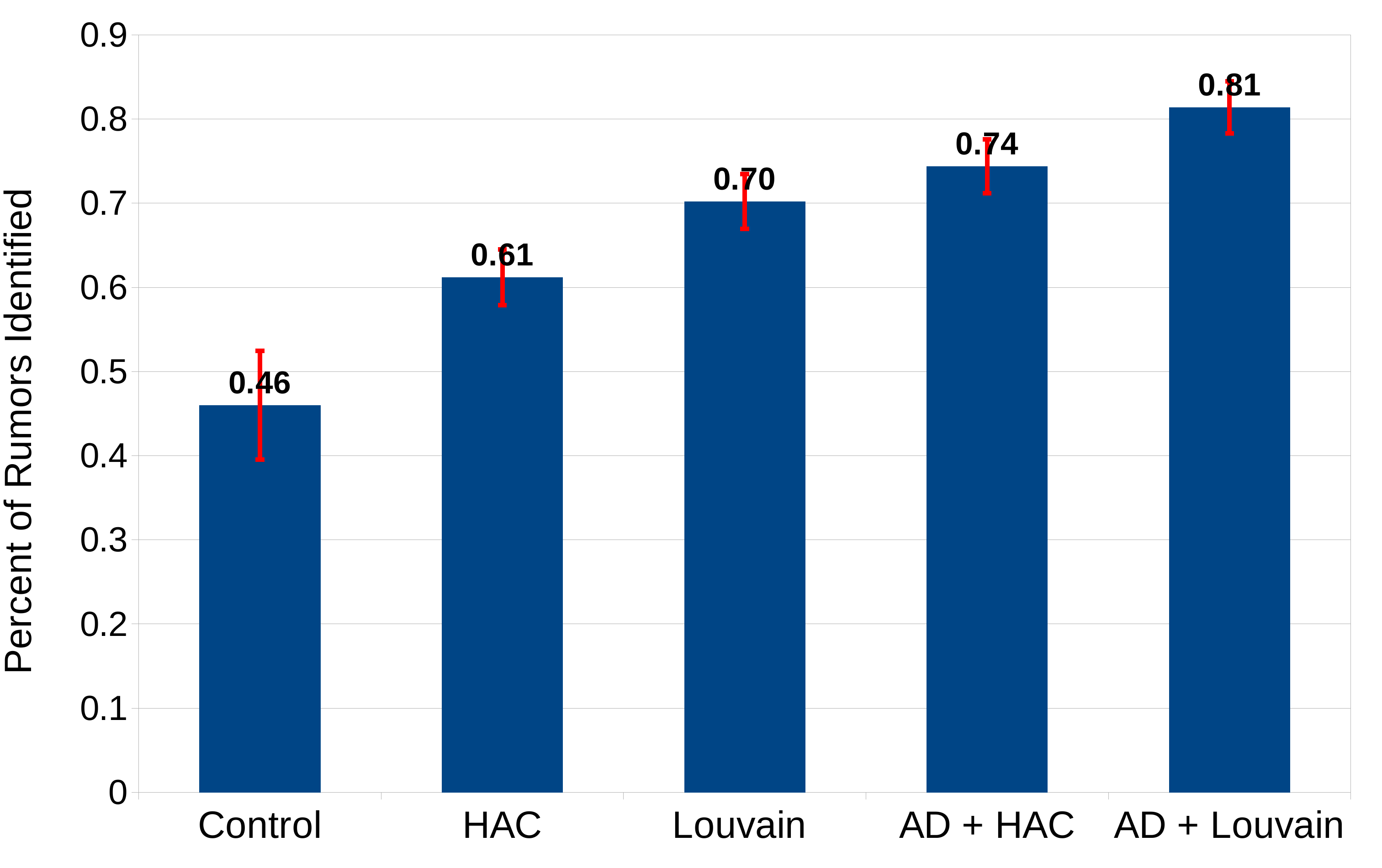}
\caption{Percentage of rumors correctly identified by subjects, using five different version of our tool.}
\label{fig:res1}
\end{figure}

After the five minutes had passed, we asked them to list the rumors they had identified. For each of the five groups, we averaged the percentage of rumors in our list that they had correctly identified. Figure \ref{fig:res1} shows the mean, and the standard deviation of each of the five groups. There was a statistically significant difference between groups as determined by one-way ANOVA ($F(4,20) = 11.283, p < .001$). There are three interesting conclusions that one can draw from these results. First, the groups using all versions of our tool outperformed the control group significantly, with the best performing group outperforming the control group by 76\%. Second, Louvain based clustering outperformed HAC, both when combine with assertion filtering (by 9\%) and when not (by 15\%). Second, assertion filtering improved the performance of both Louvain (by 16\%) and HAC clustering (by 21\%). The best performing tool was the one that combined assertion filtering with Louvain based clustering. Using that tool, the subjects were able to correctly identify 81\% of the rumors in five minutes. Though, somewhat simple, this experiment highlights the advantages of our tool, compared to other more conventional methods.


\subsubsection{Analysis 2}
Next, we quantitatively measured the quality of the clusters that are produced by HAC and Louvain clustering methods. For this analysis, we used the same dataset as the first analysis, except we did not mix in random tweets, thus we had a total of 6,825 tweets about 21 different rumors, across 3 events. We ran HAC and Louvain on the dataset and cut the generated trees so that there would be 21 clusters (same as the number of rumors). We then used the adjusted RAND index \cite{rand1971objective} and adjusted mutual information score \cite{vinh2010information}--both measures that score the similarity between two data clusterings--two compare the 21 ground-truth rumor clusters to the clusters produced by HAC and Louvain. Table \ref{tab:rand} shows the results. The results are quantitative confirmation that the Louvain is superior to HAC.

\begin{table}[!htbp]
\centering
\small
\begin{tabular}{@{}l|c|c@{}}
& Adjusted RAND & Adjusted MI \\
\hline
HAC & 0.14 & 0.19  \\
\hline
Louvain & 0.25 & 0.31    \\
\end{tabular}
\caption{Adjusted RAND and Mutual Information scores for HAC and Louvain clustering methods.}

\label{tab:rand}
\end{table}

\section{Conclusions}

In this paper, we presented a semi-automatic tool that can be used to efficiently identify stories about real-world events on Twitter. This is an important problem since Twitter and other social media platforms have become one of the main sources of news for many people. 
Given a user-specified query about an event, our tool automatically detects and clusters assertions about that event on Twitter. The system uses a Twitter speech-act classifier, in conjunction with a novel hierarchical clustering method for tweets. Instead of relying on traditional hierarchical methods which perform  poorly on tweets, our method works by first creating a similarity graph of tweets (using recent advances in Twitter NLP tools) and then applying a very fast community detection algorithm on the graph. The system is not only faster, but it also provides higher quality clusters (less noisy and more coherent), making it easier for users to quickly sort through thousands of tweets.


\small
\bibliographystyle{aaai}
\bibliography{bib}
\end{document}